\documentclass[twocolumn,showpacs,prl,floatfix,amsmath,amssymb,notitlepage,superscriptaddress]{revtex4-1}
\usepackage{epsfig}
\usepackage{subfigure}
\usepackage{amsmath}
\usepackage{color}
\usepackage{amssymb}
\usepackage{setspace}
\usepackage{graphicx}
\usepackage{dcolumn}
\usepackage{bm}
\usepackage{times}
\usepackage{enumerate}
\usepackage{footnote}
\input epsf

\newcommand{\rev}[1]{{\color{red} #1}}

\graphicspath{{figures/}}

\begin{document}

\author{Per Sebastian Skardal}
\email{persebastian.skardal@trincoll.edu} 
\affiliation{Department of Mathematics, Trinity College, Hartford, CT 06106, USA}

\author{Alex Arenas}
\affiliation{Departament d'Enginyeria Informatica i Matem\'{a}tiques, Universitat Rovira i Virgili, 43007 Tarragona, Spain}

\title{Dual control of coupled oscillator networks}

\begin{abstract}
Robust coordination and organization in large ensembles of nonlinear oscillatory units play a vital role in a wide range of natural and engineered system. The control of self-organizing network-coupled systems has recently seen significant attention, but largely in the context of modifying or augmenting existing structures. This leaves a gap in our understanding of reactive control, where and how to design direct interventions, and what such control strategies may teach us about existing structures and dynamics. Here we study reactive control of coupled oscillator networks and demonstrate dual control strategies that may be implemented interchangeably to achieve synchronization. These differing strategies take advantage of different network properties, the first directly targeting oscillator that are difficult to entrain, and the second targeting oscillators with strong influence on others. Thus, in addition to enlarge strategies for network control, the different control sets shed light on the oscillators dynamical and structural roles within the system. We close with a demonstration of the applicability of dual control of a power grid dynamics.
\end{abstract}


\maketitle

A vital benchmark for human scientific endeavor is to understand, control, and optimize, natural and man-made complex systems around us. A special challenge emerges in the context of coordinating large, network-coupled ensembles of oscillators~\cite{Strogatz2003,Pikovsky2003,Arenas2008PhysRep}. In addition to exhibiting a daunting complexity, self-organization of such systems underpin critical function in a wide range of applications such as cell circuits, cardiac pacemakers, and power grids~\cite{Prindle2012Nature,Glass1988,Rohden2012PRL}. On the one hand, the control of network dynamics, both linear and nonlinear, has seen significant attention in recent years~\cite{Cornelius2012NatComms,Yan2012PRL,Liu2013Nature,Yuan2013NatComms,Sun2013PRL,Menichetti2014PRL,Pasqualetti2014IEEE,Menara2019IEEEa,Montanari2022PNAS,Duan2022PNAS}, with attention to control and optimization of oscillator systems specifically focusing largely on the design of the underlying network structure~\cite{Hoppensteadt1999PRL,Skardal2014PRL,Fazlyab2017Automatica,Forrow2018PRX,Menara2019IEEEb,Baggio2021NatComms,Menara2022NatComms}. On the other hand, the more direct approach of reactive, or pinning, control, which directly intervenes in the system dynamics~\cite{Grigoriev1997PRL,Wang2002PhysA,Li2004IEEE,Skardal2015SciAdv}, has been explored to a lesser degree, leaving an incomplete understanding of precisely where and how to intervene in self-organizing systems to exert control. Moreover, the degree to which we may gain insight into a network's structure and/or dynamics from control implementation remains largely opaque.

In this Letter we explore reactive control of heterogeneous oscillator networks. We demonstrate that there are dual control strategies that target different sets of oscillators that may be conveniently, and interchangeably used to achieve a controlled, synchronized state. In addition to providing two different options for controlling the network dynamics, these two strategies illuminate different roles of oscillators in the networks. The first strategy, to be called row control, exerts control directly, identifying precisely those oscillators that tend to not entrain with the rest of the network on their own. The second strategy, to be called column control, acts more indirectly, targeting oscillators that do not necessarily need entrainment themselves, but rather have a strong influence on other oscillators and thus play a key role in entraining the full system. We also study the properties of the sets of oscillators that require control via the two different strategies as the overall network coupling is varied. Interestingly, as the generic properties of oscillators belonging to the column control set are largely unaffected by overall coupling, the coupling causes a transition in the properties of oscillators belonging to the row control set. Specifically, when overall coupling is small, yielding strongly disordered uncontrolled dynamics, the oscillators needing direct control are hub-like with many incoming connections. On the other hand, when overall coupling is large, yielding less disorder in the uncontrolled dynamics, the oscillators needing direct control have relatively few incoming connections. Lastly, we illustrate the effectiveness of both control strategies in an example application of a power grid where parameter heterogeneity gives rise to effective directness despite an underlying undirected topology.

In this work we consider network-coupled systems of $N$ Kuramoto oscillators~\cite{Kuramoto1984} whose dynamics are governed by
\begin{align}
\dot{\theta}_i=\omega_i+K\sum_{j=1}^NA_{ij}\sin(\theta_j-\theta_i)+f_i(t),\label{eq:01}
\end{align}
where $\theta_i$ and $\omega_i$ are the phase and natural frequency of oscillator $i$, $K$ is the global coupling strength, and the adjacency matrix $A$ encodes the structure of the underlying network, which we assume is directed, and lastly the function $f_i(t)$ represents the possible reactive control exerted onto oscillator $i$. (The choice $f_i(t)=0$ corresponds to no control imposed at all.) Here we focus on the unweighted directed case, where $A_{ij}=1$ if a link $j\to i$ exists and $A_{ij}=0$ otherwise, but our results generalize to weighted links. We focus on the (complete) synchronization of the ensemble of oscillators, i.e., we target the state where they all process with the same angular velocity,
\begin{align}
\lim_{t\to\infty}|\dot{\theta}_i(t)-\dot{\theta}_j(t)| = 0\hskip2ex\text{for all}\hskip2ex i,j=1,\dots,N.\label{eq:02}
\end{align}
It should also be noted that a common metric for the degree of synchronization in a network is Kuramoto's order parameter, given by $z=re^{i\psi}=N^{-1}\sum_{j=1}^Ne^{i\theta_j}$, whose amplitude $r$ delineates weakly and strongly clustered states ($|z|\approx0$ and $|z|\sim1$, respectively). We note, that the value of the order parameter may not easily be used to identify complete synchronization, but its asymptotic behavior may, with a completely synchronized state yielding relaxation to a fixed value, i.e., $|z(t)|\to r_\infty$ as $t\to\infty$.

Our control mechanisms begins by identifying a target, synchronized state. Specifically, we seek a target state of the form $\bm{\theta}^{\text{target}}(t)=\bm{\theta}^*+\bm{\Omega}t$, where $\bm{\theta}^*\in\mathbb{R}^N$ is a static vector of heterogeneous values and $\bm{\Omega}\in\mathbb{R}^N$ is the constant vector whose values $\Omega$ are given by the collective frequency of the synchronized system. We proceed by linearizing Eq.~(\ref{eq:01}) to obtain, in vector form,
\begin{align}
\dot{\bm{\theta}}=\bm{\omega}-KL\bm{\theta},\label{eq:03}
\end{align}
where $L$ is the Laplacian matrix whose entries are given by $L_{ij}=\sum_{l=1}^NA_{il}$ if $i=j$ and $L_{ij}=-A_{ij}$ otherwise. Inserting the target state into Eq.~(\ref{eq:03}) yields
\begin{align}
\bm{\theta}^*=K^{-1}L^{\dagger}(\bm{\omega}-\bm{\Omega}),\label{eq:04}
\end{align}
where $L^{\dagger}$ is the pseudoinverse of $L$~\cite{BenIsrael1974}. Moreover, it can be shown that the collective frequency $\Omega$ is, in general, not equal to the mean natural frequency, but in the general case of a directed network, equal to a weighted average of the natural frequencies, where the weights come from the first left singular vector $\bm{u}^1$ of $L$, i.e., $\Omega=\langle\bm{u}^1,\bm{\omega}\rangle/\langle\bm{u}^1,\bm{1}\rangle$~\cite{Skardal2016Chaos,Skardal2016PRE}.

With our synchronized target state in hand, we now seek to control the network dynamics by ensuring its stability. After choosing a control function designed to entrain oscillator towards its target state, i.e., $f_i(t)=F_i\sin(\theta_i^{\text{target}}(t)-\theta_i(t))$, where $F_i$ is the control strength used for oscillator $i$, the Jacobian at the target state is given by
\begin{align}
DF_{ij}|_{\bm{\theta}=\bm{\theta}^*}=\left\{\begin{array}{rl}-K\sum_{j\ne i}w_{ij}-F_i&\text{if }i=j\\ K w_{ij} & \text{if }i\ne j\end{array}\right.
\end{align}
where $w_{ij}=A_{ij}c_{ij}$ and $c_{ij}=\cos(\theta_j^*-\theta_i^*)$. Stability of the target state, and control of the synchronized state, is achieved by ensuring that the full spectrum of eigenvalues lies in the left-half complex plane. We note that, in the uncontrolled case (where $F_i=0$) the rows of the Jacobian sum to zero, yielding at least one trivial eigenvalue $\lambda_1=0$, corresponding to a marginal stability induced by rotational symmetry, so we seek to ensure that the real part of all eigenvalues are non-positive.

Our two different control strategies stem from the application of the Gershgorin circle theorem~\cite{Golub1996}, which can be applied to either the rows or the columns of $DF$. Following~\cite{Skardal2015SciAdv}, application of the theorem to the rows ensures that all $N$ eigenvalues lie within the union of $N$ row discs, $\bigcup D_{i}^{\text{row}}$, where the $i^{
\text{th}}$ disc $D_i^{\text{row}}$ is centered at $DF_{ii}$ and has radius $\sum_{j\ne i}|DF_{ij}|$. Beginning with no control, i.e., $F_i=0$, the $i^{\text{th}}$ discs is centered at $DF_{ii}=-K\sum_{j\ne i}w_{ij}$ and has a radius $r_i=K\sum_{j\ne i}|w_{ij}|$. This disc contains some portion in the right half complex plane, thereby admitting the possibility of an eigenvalue with positive real part, if any $w_{ij}$ in row $i$ is positive. Thus, by inspecting the rows of $DF$ we identify all oscillators that require control and for those oscillators set $F_i\ge K\sum_{j\ne i}(|w_{ij}|-w_{ij})$. We refer to this strategy as {\it row control}.

\begin{figure}[t]
\centering
\epsfig{file =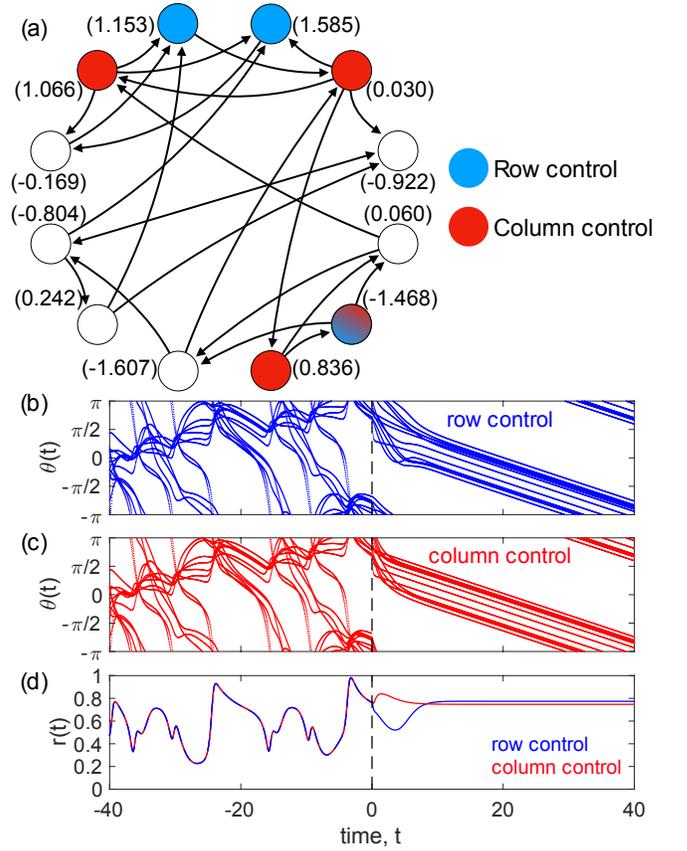, clip =,width=1.0\linewidth }
\caption{{\it Dual control strategies for complex oscillator networks.} (a) A 12-node directed network with average degree $\langle k\rangle=2$. Natural frequencies are indicated in parentheses, the coupling strength is $K=1.22$, and row- and column-controlled oscillators are filled blue and red, respectively. Time series for collection of phases under (b) row control, (c) column control, and (d) the magnitude of the order parameter $r(t)$ before ($t<0$) and after ($t\ge0$) control is applied.} \label{fig1}
\end{figure}

Our second strategy, which we refer to as {\it column control}, stems from applying the Gershgorin circle theorem to the columns of $DF$. From this perspective the eigenvalues lie within the union of column discs, $\bigcup D_j^{\text{column}}$, where $D_j^{\text{column}}$ is centered at $DF_{jj}=-K\sum_{i\ne j}w_{ji}$ and has radius $\sum_{i\ne j}|DF_{ij}|=K\sum_{i\ne j}|w_{ij}|$. It follows that oscillator $j$ requires column control if it satisfies $\sum_{i\ne j}|w_{ij}|>\sum_{i\ne j}w_{ji}$, and for each we set $F_j\ge\sum_{i\ne j}(|w_{ij}|-w_{ji})$.

We now demonstrate row and column control using a small network of $N=12$ nodes with average degree $\langle k\rangle=2$, illustrated in Fig.~\ref{fig1}(a), with randomly generated frequencies given in parentheses, and $K=1.22$. For this set of parameters we identify the sets of row (blue) and column (red) controlled oscillators, which consist of three and four oscillators, respectively, and overlap at a single oscillator. (A few practical considerations for our simulations are summarized in the SM.) In Figs.~\ref{fig1}(b) and (c) we plot the time series for the phases $\theta_i(t)$ for when row and column control are used, respectively, letting the system evolve without control for $t<0$, then turning control on for $t\ge0$. Note that, just as the control sets are different for the two strategies, the dynamics of the phases differ just after control is turned on as they relax to the synchronized state. This difference can also be seen in the dynamics of the order parameter, plotted in Fig.~\ref{fig1}(d). We emphasize that, despite the significant disorder in the phases' dynamics without control, the row and column control strategies are both effective even though their sets of oscillators are very different.

\begin{figure}[t]
\centering
\epsfig{file =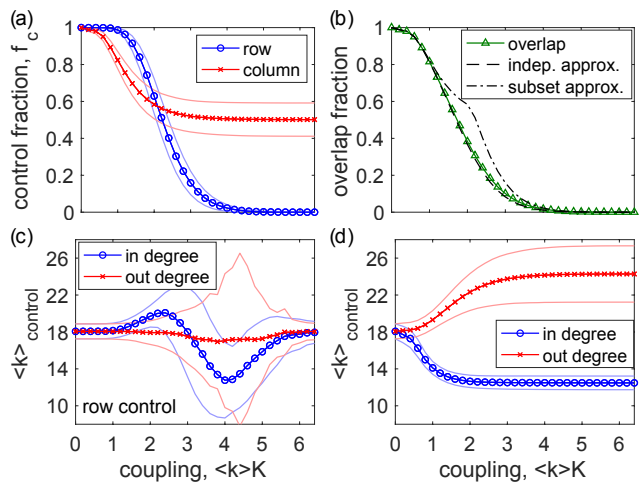, clip =,width=1.0\linewidth }
\caption{{\it Network properties of row and column control.} For an ensemble of $500$ networks with $N=200$ nodes, power-law exponent $\gamma=3$, minimum degree $k_0=9$, and normally distributed natural frequencies, (a) the mean control fraction $f_c$ for row (blue circles) and column control (red crosses) as a function of overal coupling, $\langle k\rangle K$, (b) the overlap fraction (green triangles) compared to independence and subset approximations (dashed and dot-dashed black curves). For the same ensemble of networks, the mean control in- and out-degrees $\langle k^{\text{in/out}}\rangle_{\text{control}}$ as a function of overall coupling (c) row and (d) column control. Standard deviations are indicated in (a), (c), and (d) using solid curves.} \label{fig2}
\end{figure}

We now turn to study how row and column control depend on the network structure, the structural properties of controlled oscillators in each case, and what row and column control reveal about the roles of the oscillators. We begin by plotting in Fig.~\ref{fig2}(a) the fraction of oscillators that require control, denoted the control fraction $f_c=n_c/N$, where $n_c$ is the number of oscillators in the network that require control as a function of the overall coupling $\langle k\rangle K$. Row and column control are plotted in blue circles and red crosses, respectively, and results are taken from an ensemble of $500$ networks of size $N=200$ generated with the configuration model~\cite{Bekessy1972} with power-law (uncorrelated) in- and out-degree distributions with exponent $\gamma=3$ and minimum degree $k_0=9$. For each network realization the natural frequencies are drawn from a standard normal distribution. Standard deviations are indicated by solid curves. Interestingly, these results indicate that depending on the coupling, either row or column control may be less expensive in the number of oscillators that require control. Specifically, for $\langle k\rangle K$ approximately less than or greater than $2$\rev{,} column and row control require smaller control fractions, respectively. Next, in Fig.~\ref{fig2}(b) we plot the overlap between the row and column control oscillator sets in green triangles. To better understand the overlap set, we compare this to the two simplest possible approximations bases on differing hypotheses: (i) row and column control sets are largely independent or (ii) one set tends to be nearly contained within the other. Denoting these as the {\it independence} and {\it subset} approximations, respectively, the independence approximation would indicate that the overlap set is approximately the size of the product of the row and column sets, where as the subset approximation would indicate that the overlap set is approximately equal to the minimum size of the row and column sets (note that this also serves as an upper bound on the overlap set). Plotting the independence and subset approximations in dashed and dot-dashed black, we observe that the the independence approximation is the most accurate, indicating that row and column control largely target different oscillators in the network.

We also examine the role of the controlled oscillator more closely, plotting in Figs.~\ref{fig2}(c) and (d) the mean in- and out-degrees of the controlled oscillators, denoted $\langle k^{\text{in/out}}\rangle$ as a function of overall coupling $\langle k\rangle K$ for row and column control, respectively. Beginning with row control in panel (c), we observe first that it is the in-degree of the oscillators that primarily dictate their inclusion or exclusion from the control set, while the out-degree plays a negligible role. This is perhaps not surprising due to the application of the Gershgorin circle theorem along rows, and is further verified by inspecting network ensembles with correlated in- and out-degrees, included in the Supplemental Material (SM). More interestingly, the overall coupling causes a transition in the properties of the oscillators that require control. Namely, when the coupling is relatively small (say, $\langle k\rangle K\le 3$) the oscillators that require control tend to have large in-degrees, whereas the when the coupling is relatively large (say, $\langle k\rangle K\le 3$), the oscillators that require control tend to have small in-degrees. This can be explained by the degree of disorder imposed by smaller vs larger coupling as follows. When the disorder in the dynamics is high (from small coupling) the effect of one oscillator on another via a link is destabilizing, and therefore those with more incoming links is more likely to require control to become entrained. On the other hand, when the disorder in the dynamics is low (from larger coupling) the effect of one oscillator on another via a link is stabilizing, and an oscillator with many incoming links is more likely to be entrained spontaneously without intervention. Moving to column control in panel (d) the role of in- vs out-degrees is much more clear, as column-controlled oscillators tend to balance larger out-degrees with lower in-degrees. Mathematically this can be explained: in the limit of strong coupling the target phases all take small values resulting the approximation $c_{ij}=\cos(\theta_j^*-\theta_i^*)\approx1$, in which case the application of the Gershgorin circle theorem easily yields control for column $j$ if $k_j^\text{out}>k_j^{\text{in}}$. Moreover, these local properties reveal a larger role of column controlled oscillators as those that have a relatively strong influence on other oscillators within the network. Thus, while row control achieves control by directly identifying those oscillators that require control, column control does so via the more indirect method of stabilizing those oscillators that have a strong influence on others to entrain the full system.

\begin{figure}[t]
\centering
\epsfig{file =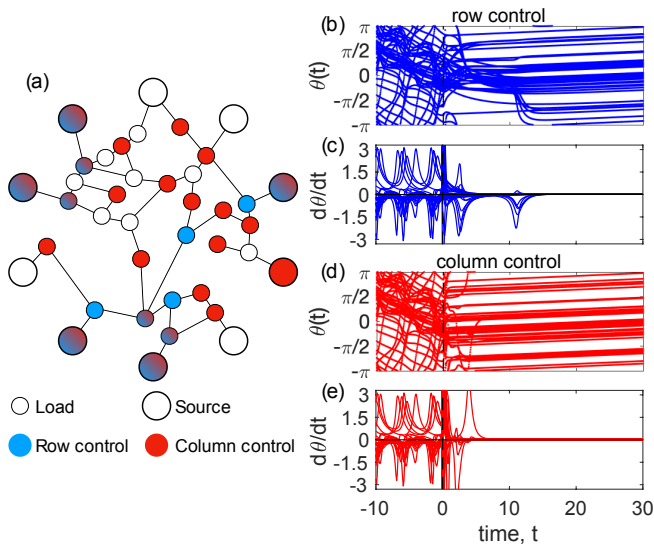, clip =,width=1.0\linewidth }
\caption{{\it Dual control of the IEEE 39 New England power grid.} (a) The 39-node IEEE New England power grid network, denoting loads and sources as small (inner) and large (outer) nodes. Natural frequencies and damping coefficients are provided in the SM. Row and column controlled oscillators are colored blue and red, respectively. Time series for collection of (b),(d) phases and (c),(e) frequencies are plotted before ($t<0$) and after ($t\ge0$) control is applied.} \label{fig3}
\end{figure}

Lastly, we present an example application using the IEEE 39 New England power grid network~\cite{Athay1979IEEE}. We first emphasize that, while a directed network structure is needed to realize different row and column strategies for an oscillator network, the presence of heterogeneous system parameters, e.g., damping, may give rise to effectively directed network structures even when the underlying topology is undirected. Noting that Kuramoto oscillator systems serve as models for a variety of types of power grids~\cite{Dorfler2012SIAM,SimpsonPorco2013Automatica,Dorfler2013PNAS,Nishikawa2015NJP} we consider the IEEE 39 New England power grid network with the following oscillator model
\begin{align}
D_i\dot{\phi}_i=p_i+K\sum_{j=1}^Na_{ij}\sin(\phi_j-\phi_i),\label{eq:06}
\end{align}
where, even for an undirected topology encoded in the entries $a_{ij}$, heterogeneity in the damping coefficients $D_i$ yield an effectively directed adjacency matrix entries $A_{ij}=a_{ij}/D_i$, as well as effective natural frequencies given by the ratio of the power and damping, $\omega_i=p_i/D_i$. In Fig.~\ref{fig3}(a) we illustrate the topology of the power grid, with small inner nodes corresponding to loads and large outer nodes corresponding to sources. Damping coefficients $D_i$ and power $p_i$ are chosen randomly, with power biased positive or negative for sources and loads, respectively. (Power and damping coefficients are given in the SM.) We then implement both row and column control, coloring row and column controlled oscillators blue and red, respectively, then plot the time series for the phases before ($t<0$) and after ($t\ge0$) control in panels (b) and (d). We also plot the time series of actual frequencies $d\theta/dt$ in panels (c) and (e). We observe that in this particular network structure, column control converges faster to synchronization, prioritizing column control versus row control. This observation is in perfect agreement with the rationale we have explained before.

To conclude, we have contributed to the understanding of reactive control in systems of heterogeneous phase oscillators' network, scrutinizing the dual strategy of row-column control. We have proven that different network topologies can be controlled optimally depending on its structure, by directly targeting oscillator that are difficult to entrain (row), or targeting oscillators with strong influence on others (column). The particular role of each oscillator in the reactive control process is revealed from its algebraic contribution on the specific control strategy. These results are proven to be applicable to real networks, and pave the way for further studies on controlling power-grid and micro-grid networks, and we speculate can even have application on neuroscience, targeting specific brain areas to (de-)synchronize neural activity.

\acknowledgments
P.S.S. acknowledges financial support from NSF grant No. MCB-2126177. A.A. acknowledges financial support from the Spanish MINECO (Grant No. PGC2018-094754-B-C2), from Generalitat de Catalunya (grant No.\ 2017SGR-896), Universitat Rovira i Virgili (grant No.\ 2019PFR-URV-B2-41), and the James S. McDonnell Foundation (grant \#220020325).

\bibliographystyle{plain}

\pagebreak
\widetext
\begin{center}
\textbf{\large Supplemental Material to ``Dual control of coupled oscillator networks''}
\end{center}
\setcounter{equation}{0}
\setcounter{figure}{0}
\setcounter{table}{0}
\setcounter{page}{1}
\makeatletter

Here we present addition results for the dual control of coupled oscillator networks. First, we consider the structural properties of row and column control in networks with in- and out-degree correlations. Next, we present numerical details for the model used and presented in Fig.~3 in the main text. Lastly, we describe some practical considerations used in control simulations in the main text.

\section{Networks Properties of row and column control: in- and out-degree correlations}

We begin with the network properties of row and column control. We investigate the control fraction $f_c$ for row and column control, the overlap between the row and column control sets, and the mean control degrees $\langle k^{\text{in/out}}\rangle$ for the row and column control sets.  As in the results presented in Fig.~2 in the main text, we consider an ensemble of 500 networks of size $N=200$ generated using the configuration model~\cite{Bekessy1972} with power-law in- and out-degree distributions with minimum degree $k_0=9$ and normally distributed natural frequencies. Here, however, rather than leaving nodal in- and out-degrees uncorrelated, we consider assortatively and diassortatively mixed networks, i.e., positive and negative correlationg between in- and out degrees, respectively. To attain these correlations, we simply match the largest in- degree with the largest or smallest out-degrees for assortative or disassortative networks, respectively, then to ensure some randomness, randomly swap the out-degrees of 2\% of the assigned nodes.

\begin{figure}[ht]
\centering
\epsfig{file =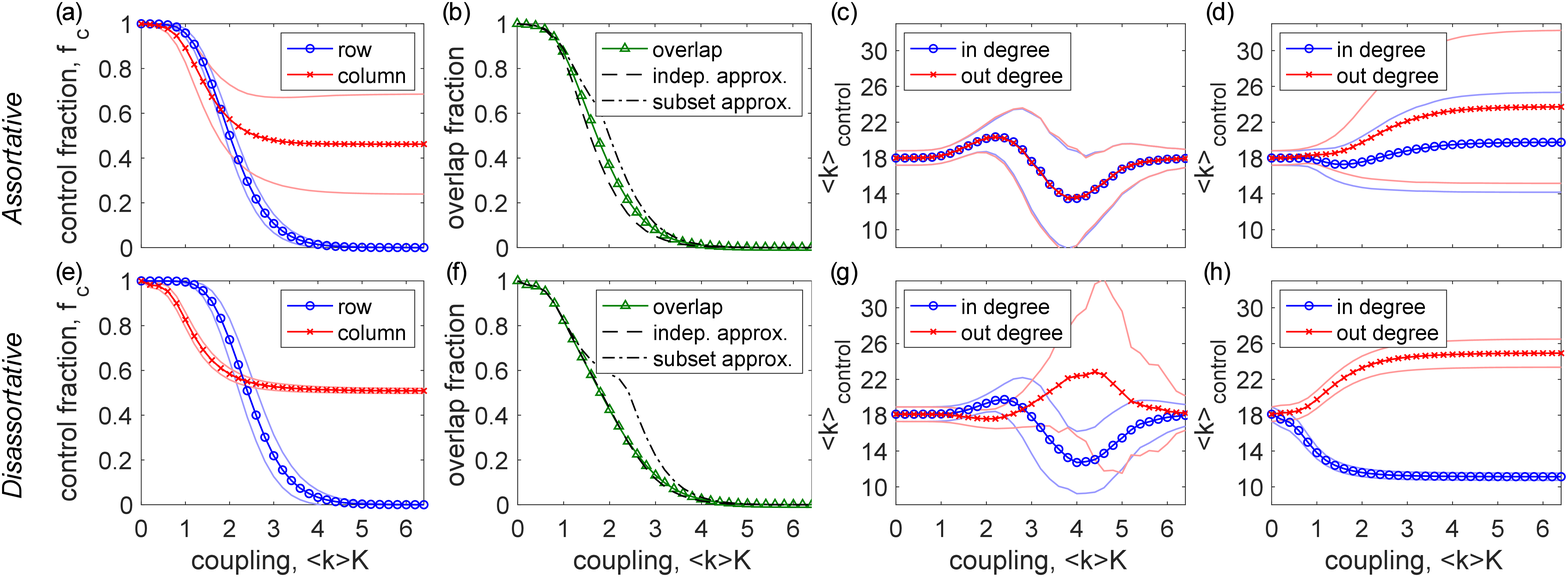, clip =,width=1.0\linewidth }
\caption{{\it Network properties of row and column control: assortative and disassortative networks.} For ensembles of $500$ networks with $N=200$ nodes, power-law exponent $\gamma=3$, minimum degree $k_0=9$, and normally distributed natural frequencies with assortative (top row) and disassortative (bottom row) in- and out-degrees, (a),(e) the mean control fraction $f_c$ for row (blue circles) and column control (red crosses) as a function of overal coupling, $\langle k\rangle K$, (b),(f) the overlap fraction (green triangles) compared to independence and subset approximations (dashed and dot-dashed black curves). For the same ensemble of networks, the mean control in- and out-degrees $\langle k^{\text{in/out}}\rangle_{\text{control}}$ as a function of overall coupling (c),(g) row and (d),(h) column control.} \label{figSM1}
\end{figure}

In Fig.~\ref{figSM1} we plot our results. For assortative networks (top row) we plot the control fraction $f_c$ for row and column control (blue circles and red crosses, respectively) in panel (a), the overlap fraction (green triangles) compared to the independence and subset approximations (dashed and dot-dashed curves, respectively) in panel (b), and the mean in- and out degrees $\langle k^{\text{in}}\rangle$ and $\langle k^{\text{out}}\rangle$ (blue circles and red crosses, respectively) for row and column control in panels (c) and (d), respectively. Similarly, for disassortive networks (bottom row) we plot the control fraction $f_c$ for row and column control (blue circles and red crosses, respectively) in panel (e), the overlap fraction (green triangles) compared to the independence and subset approximations (dashed and dot-dashed curves, respectively) in panel (f), and the mean in- and out degrees $\langle k^{\text{in}}\rangle$ and $\langle k^{\text{out}}\rangle$ (blue circles and red crosses, respectively) for row and column control in panels (g) and (h), respectively.

Importantly, the results in Fig.~\ref{figSM1} are qualitatively equivalent to those in Fig.~2 in the main text, suggesting that our general results are not specific only to uncorrelated networks. Moreover, in panels (c) and (d) we see in fact that the out-degree plays little role in shaping the row control set, as suggested in the main text.

\section{Numerical specifications of the power grid model}

\begin{figure}[ht]
\centering
\epsfig{file =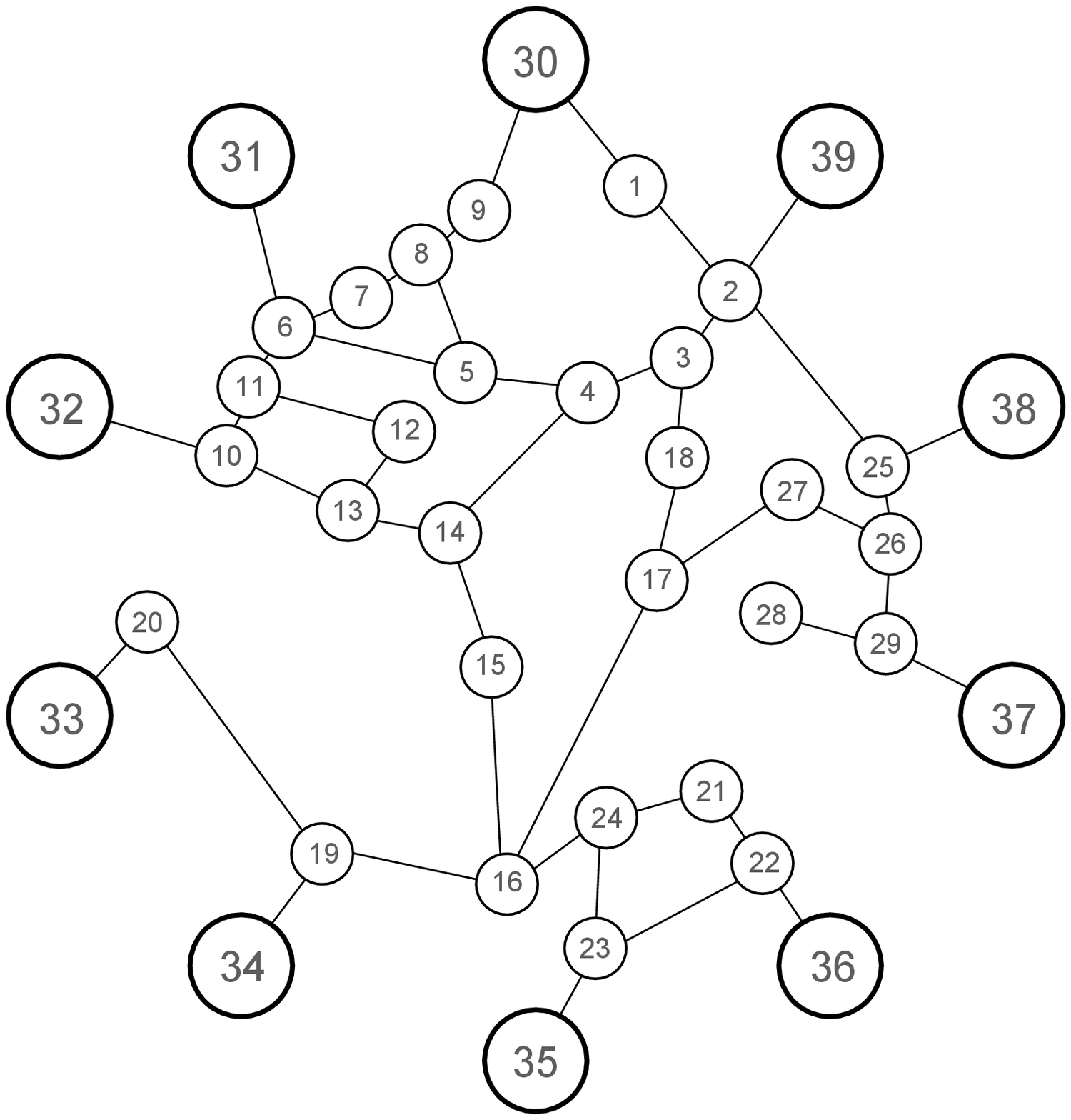, clip =,width=0.3\linewidth }
\caption{{\it IEEE 39 New England power grid topology.} Labeled indices $i=1,\dots,39$.} \label{figSM2}
\end{figure}

Next we provide the numerical parameter choices for the power grid model presented in Fig.~3 in the main text, namely, the damping coefficient, $D_i$, and power, $p_i$, for each oscillator in the network. First, in Fig.~\ref{figSM2} we illustrate the network topology of the IEEE 39 New England power grid~\cite{Athay1979IEEE} with assigned node indices, $i = 1,\dots,39$, where oscillators $i=1,\dots,29$ (inner, smaller nodes) represent loads and oscillators $i=30,\dots,39$ (outer, larger nodes) represent sources.

Each oscillator $i$ is then assigned a damping coefficient $D_i$ and power $p_i$, summarized below. Note that damping coefficients were chosen uniformly at random from the interval $[0.25,2.0]$, power was chosen randomly from one or two normal distributions to ensure that power for loads and sources were biased negative and positive, respectively. Damping and power for each oscillator are given in Table~\ref{table}.

\begin{center}
    \begin{table}[ht]
    \begin{tabular}{ | c | c | c || c | c | c |}
    \hline
    oscillator index, $i$ & damping, $D_i$ & power, $p_i$ & oscillator index, $i$ & damping, $D_i$ & power, $p_i$ \\ \hline \hline
    1 & 1.7476 & -1.0697 & 21 & 0.7782 & -0.1882 \\ \hline
    2 & 1.7524 & -1.1012 & 22 & 0.4248 & -0.1793 \\ \hline
    3 & 1.2024 & -0.0432 & 23 & 0.4054 & -0.1483 \\ \hline
    4 & 1.5625 & -0.2034 & 24 & 0.2576 & -0.1725 \\ \hline
    5 & 0.6620 & -0.2707 & 25 & 0.8430 & -1.0895 \\ \hline
    6 & 0.8039 & -0.5232 & 26 & 1.7572 & -1.1754 \\ \hline
    7 & 0.5685 & -0.4292 & 27 & 1.6242 & -0.6843 \\ \hline
    8 & 1.7742 & -1.4545 & 28 & 1.1599 & -1.1214 \\ \hline
    9 & 0.2117 & -0.1619 & 29 & 0.8356 & -0.6385 \\ \hline
    10 & 1.3095 & -0.5943 & 30 & 0.2300 & 0.4905 \\ \hline
    11 & 0.8068 & -0.3648 & 31 & 1.4356 & 2.3489 \\ \hline
    12 & 1.3422 & -0.0851 & 32 & 1.3115 & 2.2129 \\ \hline
    13 & 0.4043 & -0.2099 & 33 & 0.2642 & 0.3284 \\ \hline
    14 & 0.3241 & -0.1242 & 34 & 1.3747 & 2.2736 \\ \hline
    15 & 1.1092 & -0.8078 & 35 & 1.4981 & 2.8988 \\ \hline
    16 & 1.6324 & -0.7955 & 36 & 0.3036 & 0.3549 \\ \hline
    17 & 1.1775 & -1.0406 & 37 & 0.9786 & 1.5221 \\ \hline
    18 & 1.4898 & -0.4910 & 38 & 1.7897 & 2.8252 \\ \hline
    19 & 0.2544 & -0.2414 & 39 & 0.8235 & 1.4619 \\ \hline
    20 & 1.1897 & -0.5919 & -- & -- & -- \\
    \hline
    \end{tabular}
    \caption{Numerical parameters values for the power grid model.} \label{table}
    \end{table}
\end{center}

\section{Practical considerations used in control simulations}

Lastly, we present some practical considerations that go into control simulations, specifically those results presented in Figs.~1 and 3 in the main text. In particular, a main component of the control paradigm presented in the main text involves the identification of a target state, for which we linearize the system dynamics, i.e., moving from Eq.~(1) to Eq.(3) in the main text. This linearization is of course an approximation whose accuracy may vary depending on the network topology, natural frequencies, and coupling strength. Thus, sufficient loss of accuracy may lead to failure of the control method unless certain precautions are taken. In particular, we find that two particular practical steps may be taken to improve the success of the control mechanism.

The first practical step we take is to adjust the target state to better represent a target state obtained by solving the nonlinear system (i.e., one involving arcsines). For this we introduce a parameter $\epsilon_K$ which we use to modify the target state, specifically to use the target state
\begin{align}
\bm{\theta}^*=\frac{L^\dagger(\bm{\omega}-\bm{\Omega})}{K(1-\epsilon_K)}.
\end{align}
Here we assume that $\epsilon_K$ is positive and relatively small, so that the target state is slightly expanded, i.e., more spread out, than without $\epsilon_K$. In the main text we use $\epsilon_K=0.2$.

The second practical step we take is to add a margin of error in our identification of control oscillators depending on the rows/columns of the Jacobian. Ultimately, we seek to err on the conservative side and rather add a handful of addition oscillators to the control sets to overcome inaccuracies from approximations. For this purpose, we introduce a parameter $\epsilon_{DF}$ that is applied slightly differently in the row and column control strategies. For row control, rather than identifying entries of $DF$ where a link $j\to i$ exists and $DF_{ij}<0$, we allow for some margin of error by controlling oscillator $i$ if row $i$ contains an entry for which $j\to i$ and $DF_{ij}<\epsilon_{DF}$. For column control, we simply impose impose control on oscillator $j$ if the $j^{\text{th}}$ Gershgorin disc is within $\epsilon_{DF}$ of the right-side complex plane. Similar to $\epsilon_K$, we assume that $\epsilon_{DF}$ is positive and relatively small, so that some additional margin of error is given to ensure that eigenvalues of the Jacobian are pushed into the left half complex plane. In the main text we use $\epsilon_{DF}=0.2$.

\bibliographystyle{plain}

\end{document}